\documentclass[journal]{IEEEtran}
\usepackage{amssymb, amsmath}
\usepackage{bm}
\usepackage{color}
\usepackage{graphicx}
\usepackage{booktabs}
\usepackage{multirow}
\usepackage{threeparttable}
\usepackage{enumerate}
\usepackage[english]{babel}
\usepackage{algorithmicx}
\usepackage{algorithm}
\usepackage{algpseudocode}
\usepackage{nomencl}
\usepackage{multicol}
\usepackage{commath}
\usepackage{makecell}
\usepackage{blindtext}
\usepackage{soul}
\usepackage{comment}

\usepackage{physics}
\usepackage{caption}

\usepackage{mathrsfs}

\usepackage{pifont}

\usepackage{hyperref}
\hypersetup{
    colorlinks = false,
}

\usepackage[backend=biber,style=nature]{biblatex}
\usepackage[final]{microtype}
\usepackage{csquotes,lmodern}
\usepackage{hyperref}

\hypersetup{hidelinks}

\makenomenclature
\graphicspath{{Figures/}}

\begin{document}
\title{
Data-Driven Quantum Approximate Optimization Algorithm for Cyber-Physical Power Systems  
}
\author{\fontsize{10}{13}\selectfont Hang Jing$^{1}$,
Ye Wang$^{2,3}$*, Yan Li$^{1}$$^\dag$\\
{$^{1}$Department of Electrical  Engineering, The Pennsylvania State University, University Park, PA, 16802 USA.\\
$^{2}$Duke Quantum Center, Duke University, Durham, NC, 27708 USA.\\
$^{3}$Department of Electrical \& Computer Engineering, Duke University, Durham, NC, 27708 USA.\\
*Corresponding author: ye.wang2@duke.edu\\ $^\dag$Corresponding author: yql5925@psu.edu.
}}
% \dagger These authors contributed equally to this work.
% \thanks{H. Jing and Y. Li are with the Department of Electrical  Engineering, The Pennsylvania State University, University Park, PA, 16802 USA (e-mail: yql5925@psu.edu).}
% \thanks{Y. Wang is with the Department of Electrical \& Computer Engineering, Duke University, Durham, NC, 27708 USA.}

\captionsetup[tableExt]{font=small,labelsep=newline,textfont=sc,justification=centering}

\captionsetup[table]{font=small,labelsep=newline,textfont=sc,justification=centering}

\captionsetup[figure]{font=small,labelsep=period}

\captionsetup[figureExt]{font=small,labelsep=period}

\twocolumn[
\begin{@twocolumnfalse}
\maketitle
\begin{abstract}
\noindent Quantum technology provides a ground-breaking methodology to tackle  challenging computational issues in power systems, especially for Distributed Energy Resources (DERs) dominant cyber-physical systems that have been widely developed to  promote energy sustainability.
The systems' maximum power or data sections  are essential for 
monitoring, operation, and control, 
while high computational effort is required. 
Quantum Approximate Optimization Algorithm (QAOA) provides a promising means to  search  for these sections by leveraging quantum resources.
However, its performance highly relies on the critical parameters, especially for weighted graphs.
We present a data-driven QAOA,
which transfers quasi-optimal parameters between weighted graphs based on the normalized graph density, and verify the strategy with $39,774$ instances. Without parameter optimization, our  data-driven QAOA is comparable with the Goemans-Williamson algorithm.
This work advances QAOA and pilots  the practical application of quantum technique to power systems in noisy  intermediate-scale  quantum devices, heralding its  next-generation computation   in the quantum era. 
\end{abstract}

% \begin{keywords}
% \hl{Quantum  Approximate  Optimization  Algorithm (QAOA), cyber-physical power systems, parameter transfer strategy,  maximum section, Max-Cut, Hamiltonian, adiabatic evolution.}
% \end{keywords}
\end{@twocolumnfalse}]

\section{Introduction }
Quantum technology is emerging as a new hope to address challenging computational tasks in power systems, including quantum chemistry simulation for new type batteries~\cite{aspuru2005simulated, hastings2014improving,o2016scalable}, efficient power system analysis by solving linear systems of equations~\cite{harrow2009quantum,eskandarpour2021experimental,zhou2021quantum,eskandarpour2020quantum,9423668}, forecasting highly chaotic systems~\cite{PhysRevA.101.010301}, scheduling and dispatching power grids~\cite{giani2021quantum}, unit commitment~\cite{koretsky2021adapting}, optimal reconfiguration of distribution grids~\cite{silva2021qubo}, etc.
However, the existing algorithms require substantial quantum resources, limiting their near-term utilization on noisy intermediate-scale quantum (NISQ) devices~\cite{preskill2018quantum}.
Even though specific instances of quantum algorithms have been demonstrated on various quantum processors with tens of qubits~\cite{harrigan2021quantum,pagano2020quantum,graham2021demonstration},  practical applications to address power system problems will still require further advances in algorithmic design.

\begin{figure*}[ht]
	\centering{\includegraphics[width=\textwidth]{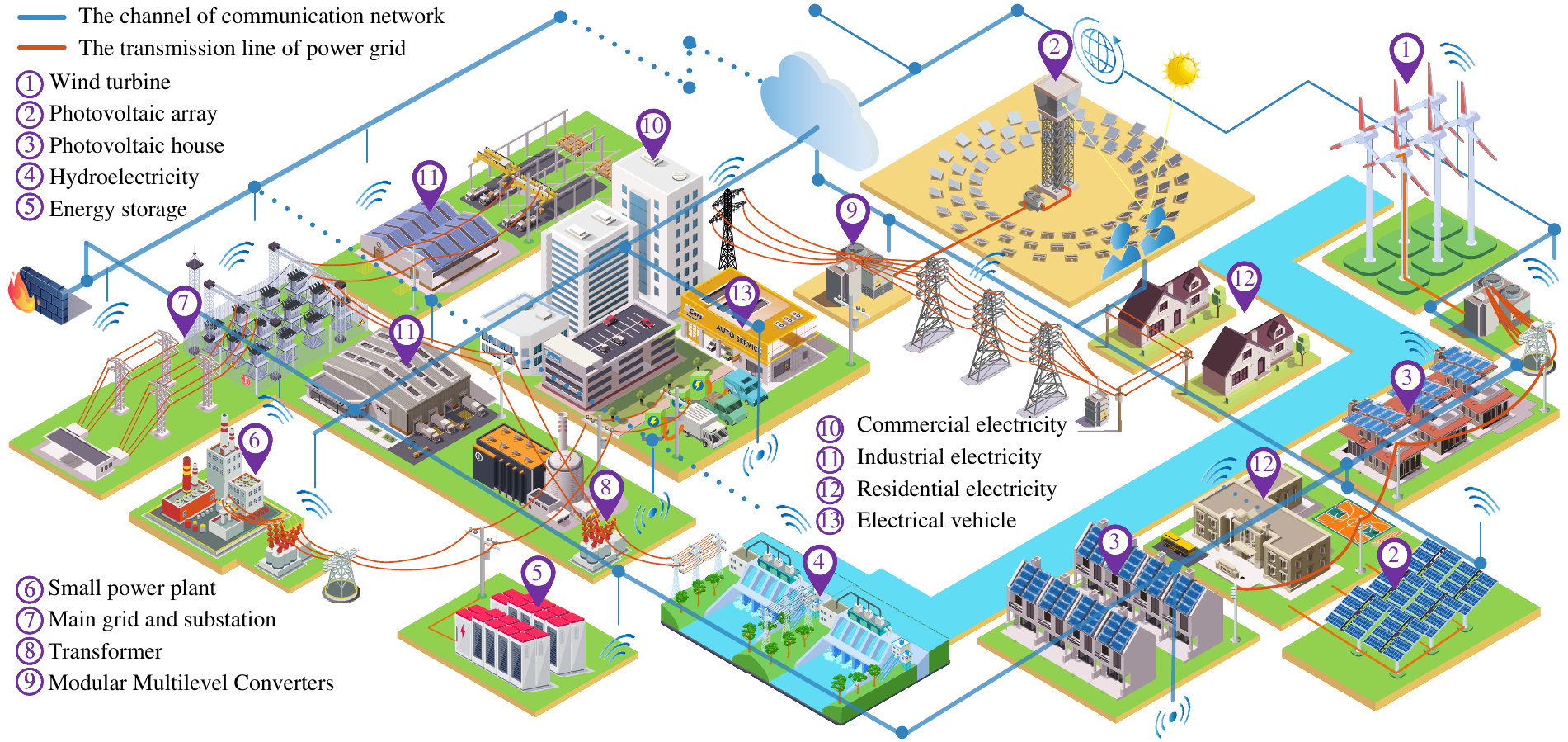}}
	\caption{Illustration of the cyber-physical power system. It includes physical layer and cyber layer. The physical layer is DER dominant power grid.
    The cyber  layer is used for the communication among DERs and the control center for the system's operation and control.
	}
	\label{fig_cyber_physical_system}
\end{figure*}

In power systems, one emerging quantum application is to analyze the Distributed Energy Resources(DERs) dominant power system, which provides a potent solution to seek an edge toward energy sustainability.
We illustrate a typical DER dominant power system in 
Fig.~\ref{fig_cyber_physical_system}, where the  physical system is energized by DERs and the cyber-layer enables the communication among DERs for  coordination and control. 

To improve the resiliency of the cyber-physical system, it is critical to efficiently obtain the maximum sections of power energy in the physical layer and data traffic in the cyber-layer~\cite{shariatzadeh2014real,Jing2022interoperation,newman2006modularity}. 
Mathematically, finding the maximum section of power energy or data traffic is to solve the Max-Cut problem, an NP-hard problem~\cite{GAREY1976237}. 
Therefore people implement classical approximation algorithms~\cite{goemans1995improved,mei2017solving,shao2018simple,yao2019experimental} to address the Max-Cut problem in practical applications.
However, for specific instances, classical algorithms can only guarantee an approximation ratio of $0.878$~\cite{goemans1995improved,khot2007optimal}.

The Quantum Approximate Optimization Algorithm (QAOA), a hybrid quantum-classical algorithm, is expected to obtain better approximate solutions than any existing classical algorithms~\cite{farhi2014quantum,basso2021quantum}. QAOA utilizes a classical computer trains the parameters for quantum circuit~\cite{farhi2014quantum}.
The parameterized quantum circuit approximates the adiabatic evolution from an initial Hamiltonian, whose ground energy state is easy to prepare, to a final Hamiltonian, whose ground energy state encodes the solution of the Max-Cut problem. 
With an ideal approximation, people expect to obtain the exact solution of the Max-Cut problem with high probabilities \cite{farhi2000quantum}. 
Consequently, the parameters involved in the quantum circuit play an essential role in getting high-quality approximations~\cite{zhou2020quantum,wang2021quantum,yao2020policy}.
However, how to efficiently obtain appropriate parameters is still an open question. 

This work presents a data-driven QAOA with  parameter transfer  strategy to tackle the challenging issue of efficiently obtaining  appropriate parameters. 
Based on the normalized graph densities~\cite{tokuyama2007algorithms}, the quasi-optimal parameters for seed (or existing) graphs  are  transferred  to target (or new) graphs, 
which can be directly applied to obtain the Max-Cut or as an initial guess for further optimizing iterations. 
The  transfer  strategy is extendable and designed for generic weighted graph, which is extension studies of regular graphs~\cite{brandao,galda}. 
We numerically justify the strategy by the QAOA's performance on $1710$ random instances under transferred parameters and the ones after optimization.
Afterwards, we perform the data-driven QAOA in practical cyber-physical power systems to get the maximum  sections.
Simulations on $996$ case studies %weighted graphs 
have validated that the data-driven QAOA can efficiently obtain  comparable results to Goemans-Williamson(GW) algorithm, a famous classical algorithm~\cite{goemans1995improved}.
Additionally, simulation results show that the near-term achievable noise of quantum processor is negligible to the data-driven QAOA's performance.
This work would significantly reduce the computational effort for training QAOA parameters, advance the development of QAOA, and highly promote its wide applications for solving  engineering problems. This is the first practical quantum application that is feasible shortly in the NISQ era to address problems in power systems.

\section{Maximum Sections Problem Formulation}

The maximum power or data section is a segment along which the power delivery in the power grid or data traffic in the communication network reaches the maximum value.
It is critical to efficiently obtain the maximum sections because of three reasons.
First, the maximum power energy section offers a  cost-effective way to monitor the dynamics and  power delivery capability of the physical system considering the fluctuations of DERs and/or the frequent changes of system topology due to the join or removal of subsystems (e.g., microgrids)~\cite{shariatzadeh2014real}.
Second,  the maximum power energy sections cast light on    the dynamic system's  control and operations.
Dispatchable DERs can be coordinated for reducing the electric power over the maximum section to avoid system collapse \cite{Jing2022interoperation}.
Third, the maximum  data traffic sections provide an insight into enhancing the overall system's resilience through strategically designing and managing the communication network~\cite{newman2006modularity}, e.g., packet routing and traffic control.

Mathematically,  finding  the  maximum section is to solve a Max-Cut problem of a weighted graph  $G = (V, E)$, where $\abs{V}=n$ is the vertex number, $\abs{E}=m$ is the edge number, and  $w_{ij}$ represents the normalized weight of the edge $\left \langle i,j\right \rangle\in E$, where ${\rm{Max}}({w_{ij}}) = 1$.
The Max-Cut solutions are identical before and after normalization.
The edge weight is obtained via power flow calculation for the physical layer and means the data traffic in the cyber layer. Modeling details are provided in the Methods Section F.

The objective is to find a subset  $S\subset V$ that maximizes $\sum_{i \in S, j \notin S}^{}{w_{ij}}$ for cyber or physical layers, respectively. Suppose an $n$-bit string $ Z=  z_1\cdots  z_i\cdots  z_j \cdots  z_n \in \{-1,1\}^{n}$
can  denote the status of vertices $V$, showing each bit $ z_i$ will be equal to $1$ if the $i^{\text {th}}$ vertex is in the subset $S$, otherwise be $-1$. We can exhibit the partition of vertices for obtaining the maximum section. Thus, the classical cost function of the Max-Cut problem can  be defined as,
\begin{equation} \label{eq_objfunc} 
%\max_{Z\in \{-1,1\}^{n}} 
C( Z)=\sum_{ {\left \langle i,j\right \rangle} \in E}^{} w_{ij}\frac {1- z_i  z_j} {2}=\sum_{{\left \langle i,j\right \rangle} \in E}^{} w_{ij} C_{ij}( Z),
\end{equation}
where $C_{ij}( Z)$ represents  the contribution of $w_{ij}$ to the cost function.
The Max-Cut problem translates into finding the $n$-bit string $Z$ to maximize the cost function $C(Z)$.
Given the $n$-bit string $Z$, we define the approximation ratio to be $C(Z)/C(Z_{\rm Max-Cut})$, where $Z_{\rm Max-Cut}$ is the exact Max-Cut solution.
The goal of approximate algorithms is to find the solution with a high approximation-ratio.

\begin{figure*}[ht]
	\includegraphics[width=\textwidth]{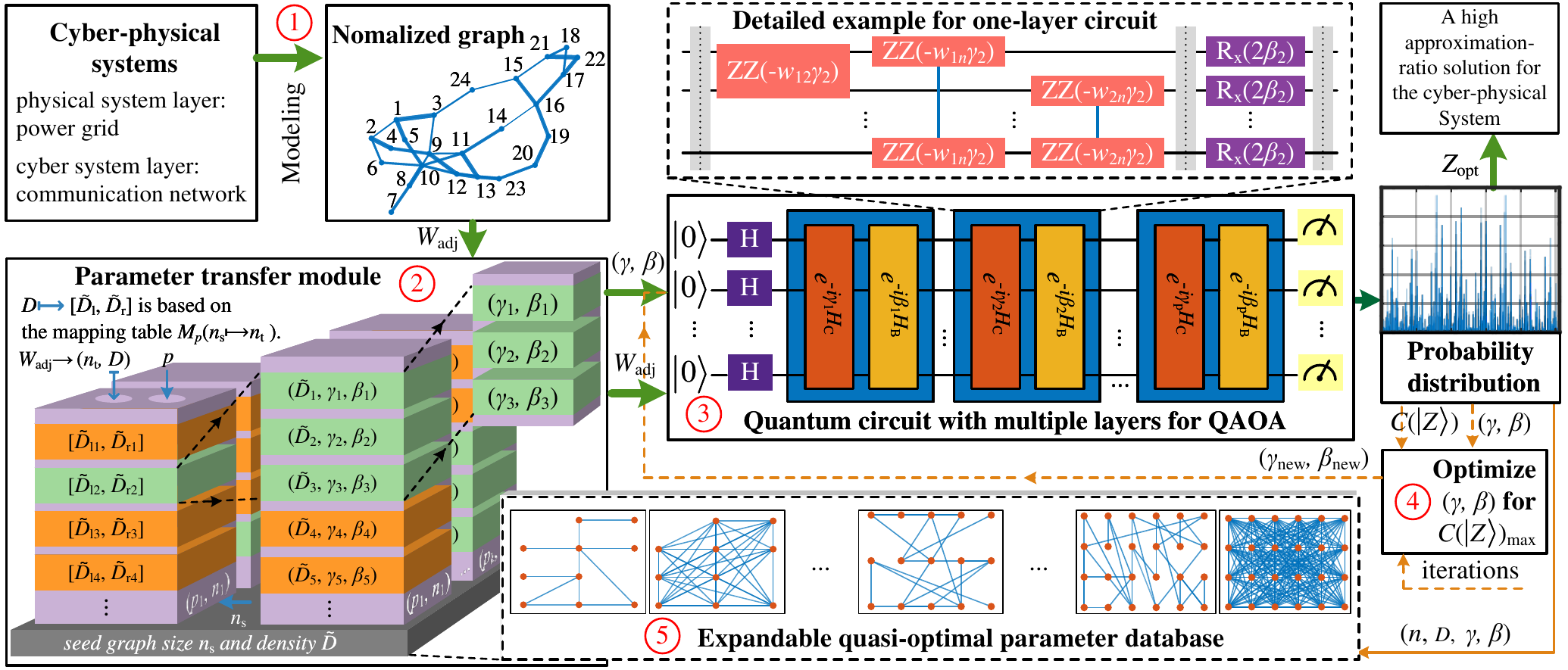}
	\caption{Schematic of the proposed data-driven QAOA for obtaining the maximum sections of cyber-physical power systems. The key idea includes the following five steps: 
	\ding{172} By modeling the \emph{Cyber-physical systems} into two normalized weighted graphs, we compute the \emph{Normalized graph} density  from adjacency matrix $W_{\rm{adj}}$. 
	\ding{173}  In the \emph{Parameter transfer module}, we first determine the seed graph size $n_{\rm{s}}$ and layer number $p$.
	Then, according to the mapping table $M_{p}{(n_{\rm{s}} \mapsto n_{\rm{t}})}$ in Extended Fig. \ref{fig_angles_index}, we can obtain the quasi-optimal parameters $(\gamma,\beta)$ from seed graphs, whose normalized densities can be expressed by an interval $[\tilde{D}_{l},\tilde{D}_{r}]$.
	\ding{174}
	The transferred parameters $(\gamma,\beta)$ are  directly passed to the \emph{Quantum circuit with multiple layers for QAOA}. By measurement, it generates the probability distribution $|\alpha_k|^2$ in (\ref{eq_quantum_cost_function}), from which we can obtain $C(\ket{Z})$ and select a high approximation-ratio  solution. 
	\ding{175} If a better performance is desired, we will further \emph{Optimize $(\gamma,\beta)$ for $C(\ket{Z})_{\rm{max}}$}.
	This step is optional.
	\ding{176}
	The obtained pair $(n,D,\gamma,\beta)$ can be used to develop an \emph{Expandable quasi-optimal parameter database} to provide quasi-optimal parameter for new target graphs. 
	If step \ding{175} is  performed, by obtaining the optimized parameters, we can store a new pair $(n_{\rm{t}},D,\gamma_{\rm{new}},\beta_{\rm{new}})$ in a new  mapping table for target graphs;
	otherwise, we can add one new entry $C(\ket{Z})$ to the original mapping table, as exampled in Extended Fig. \ref{fig_angles_index}. This step is also optional.
  	}
	\label{fig_system}
\end{figure*}

\section{QAOA for Max-cut Problem}
On quantum computers, we use $n$ quantum bits (qubits) $\ket{Z}=\ket{z_1\cdots z_i\cdots z_j\cdots z_n}$ to represent the status of $n$ vertexes. 
Each qubit $\ket{z_i}$ can be a superposition of quantum states $\ket{0}$ and $\ket{1}$, denoted as $\ket{z_i} = a_i \ket{0}+b_i \ket{1}$, where $\ket{0}$ and $\ket{1}$ are the eigenstates of the Pauli-Z operator $\sigma^z$ with the eigenvalues of $1$ and $-1$ respectively.
When we measure the qubit in the computational basis, which is $z$ basis, according to the quantum mechanics, the qubit could collapse to the state $\ket{0}$ with probability of $|a_i|^2$ and the state $\ket{1}$ with probability of $|b_i|^2$.
Therefore, unlike classical computers, the measurement results could vary even though the qubit is identical at each execution.
If we consider that measuring $\ket{0}$ represents $z_i = 1$ and measuring $\ket{1}$ represents $z_i = -1$, we can obtain various $n$-bit strings $ Z=  z_1\cdots  z_i\cdots  z_j \cdots  z_n \in \{-1,1\}^{n}$ and calculate $C(Z)$ in (\ref{eq_objfunc}) after every single quantum computer execution.

On the other hand, we could obtain the deterministic $n$-bit string $Z_k$ out of the measurements on $2^n$ $n$-qubit eigenstates in the computational basis, denoted as $\ket{Z_k}$ with $\ket{z_{k,i}} = \ket{0}$ or $\ket{1}$ and $z_{k,i} = \bra{z_{k,i}}\sigma_i^z\ket{z_{k,i}}$.
Therefore, we can have
\begin{eqnarray}
  C(Z_k) &=& \sum_{ {\left \langle i,j\right \rangle} \in E}^{} w_{ij}\frac {1- z_{k,i}  z_{k,j}} {2} \nonumber\\
  &=& \sum_{ {\left \langle i,j\right \rangle} \in E}^{} w_{ij}\frac {1- \bra{z_{k,i}}\sigma_i^z\ket{z_{k,i}}  \bra{z_{k,j}}\sigma_i^z\ket{z_{k,j}}} {2} \nonumber\\
  &=& \bra{Z_k}H_{\rm{C}}\ket{Z_k} \nonumber \\
  &\equiv& C(\ket{Z_k}),
\end{eqnarray}
where
\begin{equation}
    \label{eq_final_Ham}
    H_{\rm{C}} =\sum_{ {\left \langle i,j\right \rangle} \in E}^{} w_{ij} \frac{I-\sigma_i^z \sigma_j^z}{2}.
\end{equation}
We consider $C(\ket{Z}) = \bra{Z}H_{\rm{C}}\ket{Z}$ as the quantum analog of $C(Z)$. 
Then $2^n$ classical cost functions $C(Z_k)$ are one-on-one mapped to $2^n$ quantum cost functions $C(\ket{Z_k})$.
The Max-Cut problem translates into finding the quantum state $\ket{Z_k}$ to maximize the cost function $C(\ket{Z_k})$.

The $2^n$ $\ket{Z_k}$ states form the complete basis of the $2^n$ Hilbert space for $n$-bit quantum states. 
Therefore we can decompose an arbitrary state $\ket{Z}$ into a linear combination of $\ket{Z_k}$, denoted as $\ket{Z}=\sum_{k=1}^{2^n} \alpha_k \ket{Z_k}$ with $\sum_{k=1}^{2^n} |\alpha_k|^2 = 1$.
The quantum cost function of an arbitrary state $\ket{Z}$ can be written as
\begin{eqnarray}
\label{eq_quantum_cost_function}
    C(\ket{Z})&=& \bra{Z} H_{\rm{C}} \ket{Z} \nonumber\\
    &=& (\sum_{k=1}^{2^n} \alpha_k^* \bra{Z_k}) H_{\rm{C}} (\sum_{k=1}^{2^n} \alpha_k \ket{Z_k}) \nonumber\\
    &=& \sum_{k=1}^{2^n} |\alpha_k|^2 C(\ket{Z_k}).
\end{eqnarray}
Since $C(\ket{Z}) = \bra{Z} H_{\rm{C}} \ket{Z}	\geq0$, we have $\max C(\ket{Z}) = \max_{k=1}^{2^n} C(\ket{Z_k})$.
Notably, in quantum mechanics, $C(\ket{Z}) = \bra{Z} H_{\rm{C}} \ket{Z}$ is the expectation value of system energy for a quantum system described by Hamiltonian $H_{\rm{C}}$.
The Max-Cut problem translates into finding the maximum energy state for the quantum system described by Hamiltonian $H_{\rm{C}}$.

QAOA utilizes a quantum circuit running on the quantum computer to approximate an adiabatic evolution from the maximum energy state of an initial Hamiltonian, $H_{\rm{B}}$, to the maximum energy state of the final Hamiltonian, $H_{\rm{C}}$.
For the Max-Cut problem, we particular define the $H_{\rm{B}}$ as
\begin{equation} 
\label{eq_intial_state}
H_{\rm{B}} = \sum_{j=1}^n \sigma_j^x.
\end{equation}
According to adiabatic theorem~\cite{born1928beweis}, with an ideal approximation, we expect to obtain the maximum energy state of $H_{\rm{C}}$, which leads to the exact Max-Cut solution, with a high probability.

To implement QAOA on quantum computers, we first prepare the maximum energy state of $H_{\rm{B}}$, ${\ket{+}}^{\otimes n}$, as the initial state for the quantum circuit.
Then we run the quantum circuit with $2p$ trainable parameters ${{\gamma}}=\left(\gamma_1,\gamma_2,\ldots,\gamma_{p}\right)$ and ${{\beta}}=\left(\beta_1,\beta_2,\ldots,\beta_{p}\right)$ to approximate the $p$-step Trotter expansion of the adiabatic evolution. 
We measure the output state, obtain the classical $n$-bit string, and estimate the quantum cost function using (\ref{eq_quantum_cost_function}) with multiple executions.
After that, we use the classical-quantum hybrid optimizer iterating $2p$ parameters to maximize the quantum cost function.
Ideally, when $p$ tends to infinity, the probability of obtaining the exact Max-Cut solution will tend to be 1.
Even with a finite $p$, measuring the final state $\ket{Z}$ of the optimized circuit could generate high approximation-ratio solutions.
More details are discussed in the Methods Section A.

\section{ Data-Driven QAOA}

People have studied QAOA's efficiency and accuracy in regular graphs with constant circuit depth \cite{farhi2014quantum,galda,wurtz2021maxcut,brandao,guerreschi2017practical}.
On the other side, the performance of QAOA on generic weighted graphs is an open question and challenging to estimate rigorously~\cite{barak2021classical}.
Heuristic strategies show potentials to find quasi-optimal parameters with high approximation-ratio solutions, a claim backed by numerical evidences~\cite{zhou2020quantum,brandao,galda}. 
However, researchers have not exhaustively explore the heuristic strategies on generic weighted graphs~\cite{barak2021classical}.

Our data-driven QAOA for generic  weighted  graphs can provide a high approximation-ratio solution without parameter optimization to  avoid expensive computational effort. 
The data-driven QAOA is based on normalized weighted graph density $D$~\cite{tokuyama2007algorithms}, which is defined as,
\begin{equation} \label{eq_average_degree} 
D = \sum_{{\left \langle i,j\right \rangle} \in E}^{}{\frac{2w_{ij}}{n(n-1)}},
\end{equation}

The data-driven QAOA includes five steps as  shown in Fig. \ref{fig_system} and introduced as follows, where an innovative parameter transfer strategy is the key idea. 

\textbf{Step 1:} Formulate the search of maximum section  to the Max-Cut problem of the  normalized weighted graph and calculate its $D$.

\textbf{Step 2:} 
Obtain the quasi-optimal  parameters $(\gamma,\beta)$ based on $D$ via the parameter transfer strategy, and then
pass these parameters to the quantum processors. 
The transfer strategy  works for generic weighted graphs. 

\textbf{Step 3:} Construct the quantum circuit with the adjacency matrix $W_{\rm{adj}}$ and parameters $(\gamma,\beta)$, and  run it in a quantum processor. 
Then, measure the output state of the quantum circuit to get the probability distribution and calculate the cost function value.

\textbf{Step 4:} 
Optimize the parameters by the classical optimizer for a better result when necessary. Step $4$ is optional. 

\textbf{Step 5:}
Expand the database by adding more pairs denoted by $(n,D,\gamma,\beta)$ from verified cases, to provide more quasi-optimal parameters. Step $5$ is optional.

Decent initial guesses obtained in Step 2 can also help to handle noise-free barren plateaus, which are linked  to random parameter initialization~\cite{cerezo2021cost}. 
These initial guesses also significantly reduce the iterations between the classical optimizer and the quantum processor, saving the running time of the algorithm.

\section{Parameter Transfer Strategy}

\begin{figure*}[ht]
	\centering{\includegraphics[width=\textwidth]{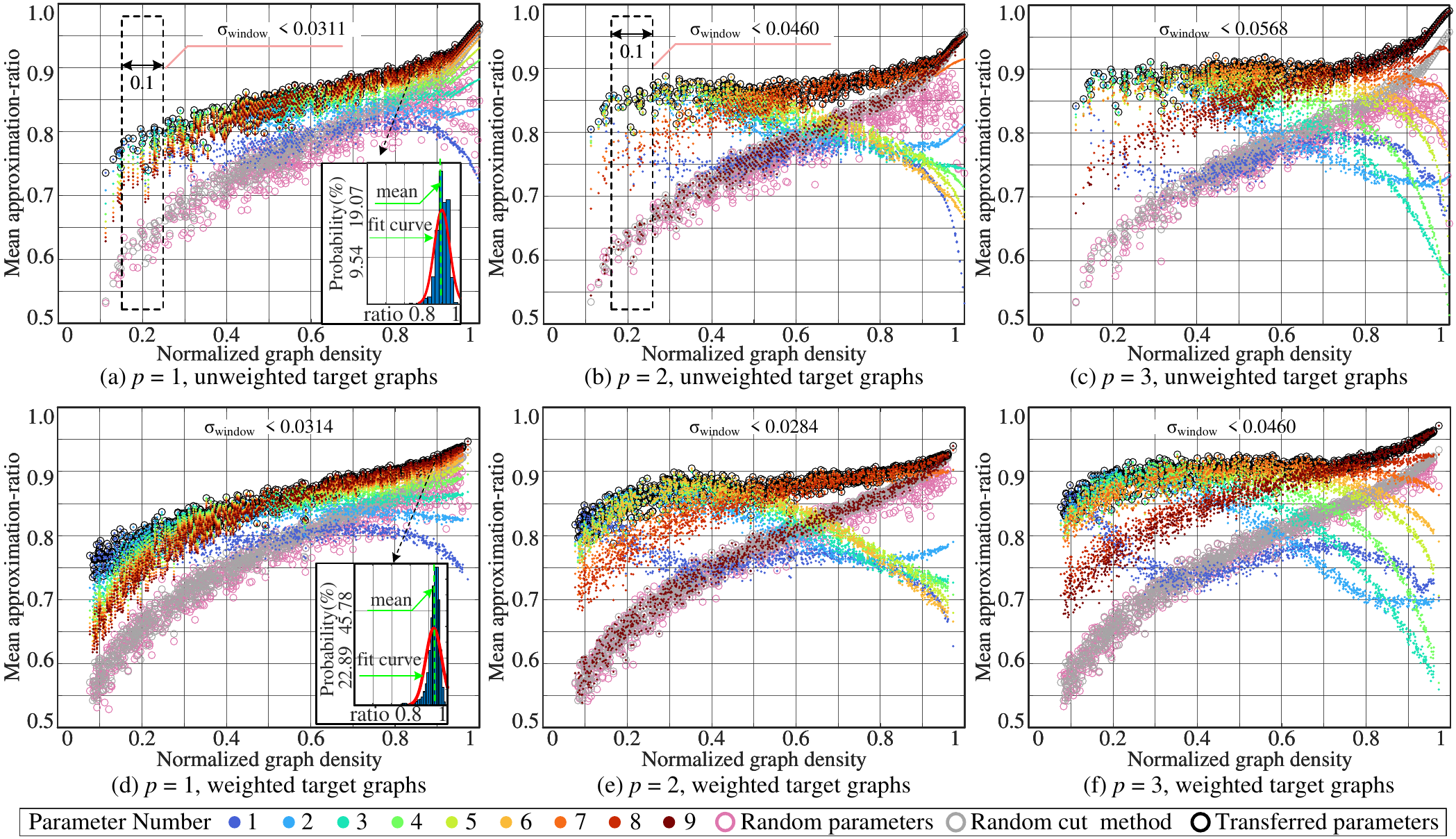}}
	\caption{The mean  approximation-ratio of randomly generated target graphs under parameters of seed graphs with $n_{\rm{s1}}=10$. The mean approximation-ratios are computed by the probability distribution, with details given in (\ref{eq_quantum_cost_function}) and the Methods Section B. The probability with respect to approximation-ratio is fitted by normal distribution. The $\sigma_{\rm{window}}$ is the standard deviation of the scatters with the same color in the $0.1$ scan window regarding to $D$. 
	}
	\label{fig_meanValue_before_cancel_random}
\end{figure*}

\begin{figure*}[ht]
	\centering{\includegraphics[width=\textwidth]{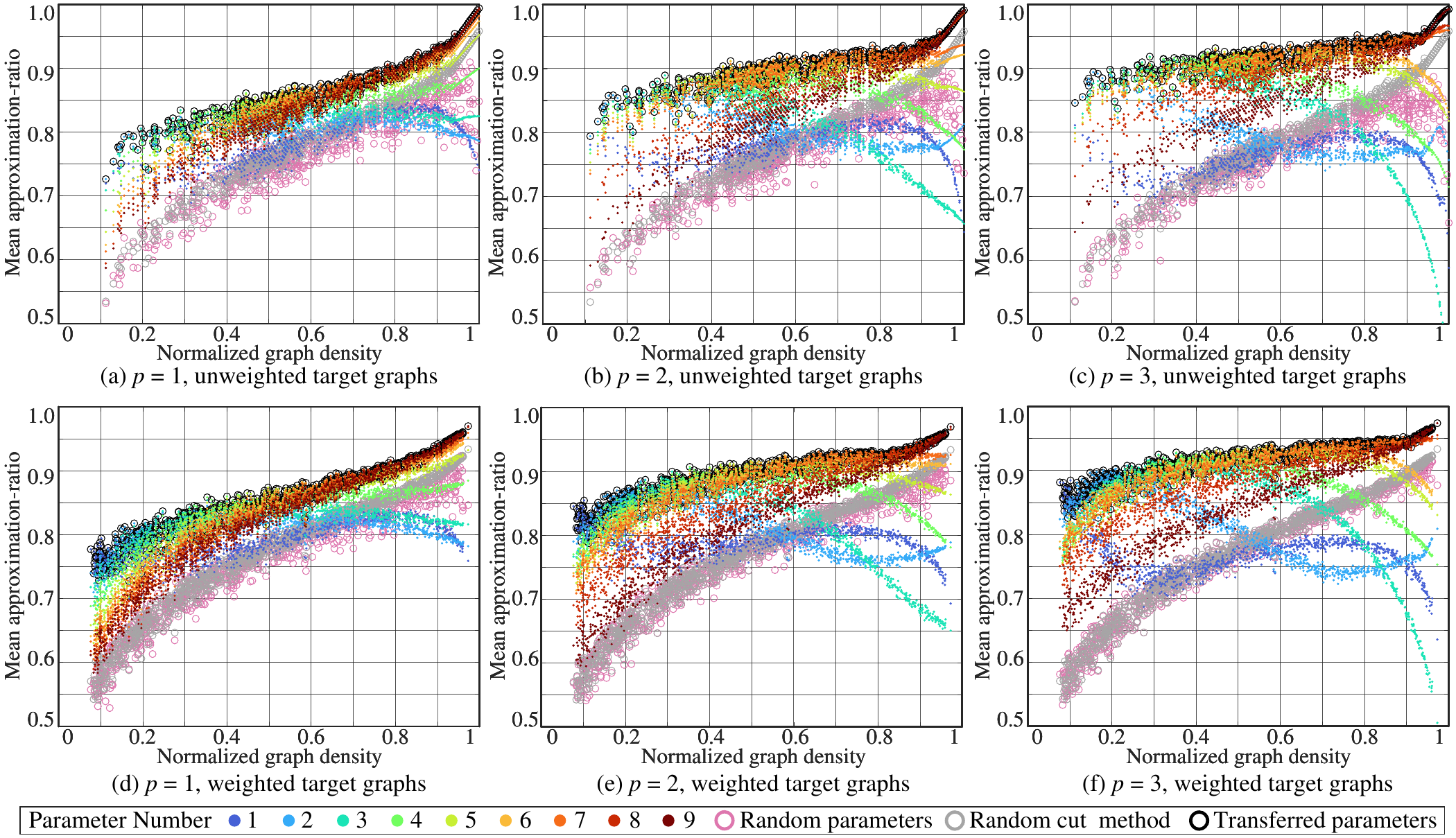}}
	\caption{The mean  approximation-ratio  of randomly generated target graphs under parameters of seed graphs with $n_{\rm{s2}}=24$. }
	\label{fig_meanValue_24nodes_before_cancel_random}
\end{figure*}

\begin{figure*}[ht]
	\centering{\includegraphics[width=\textwidth]{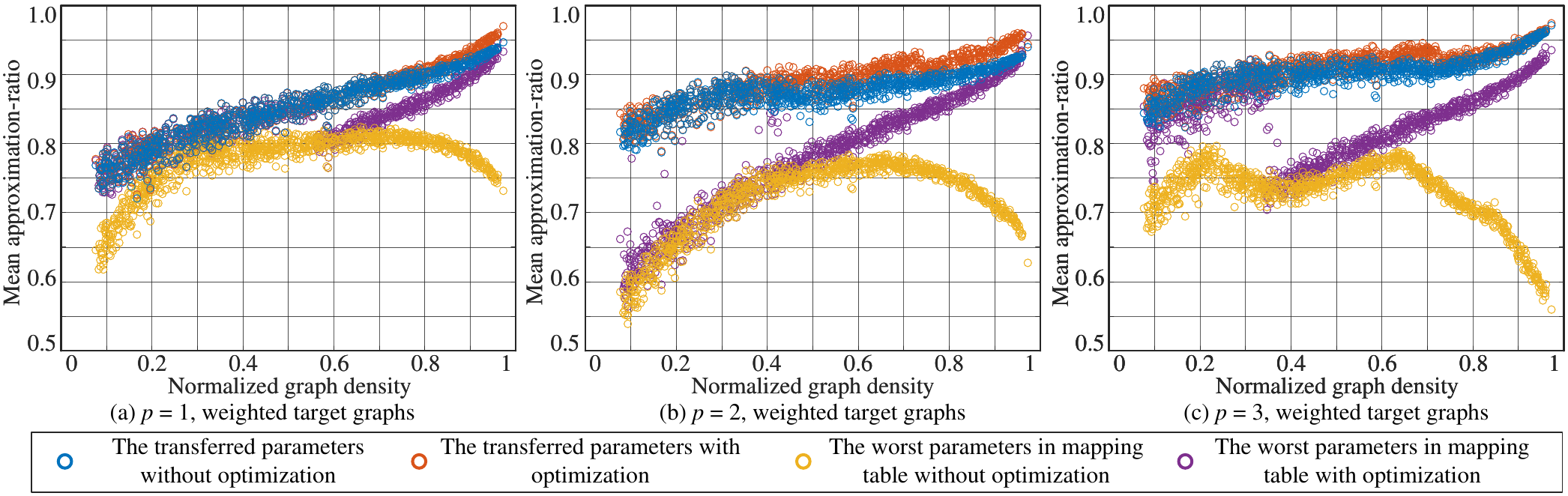}}
	\caption{The mean  approximation-ratio  of QAOA results after optimization.}
	\label{fig_meanValue_iteration}
\end{figure*}

The essential idea of the  data-driven QAOA in Fig.~\ref{fig_system} is the parameter transfer strategy, as summarized in the Step $2$. 
It includes the following three substeps, which highly improves the  effectiveness of transfer. 

\textbf{Substep 1: Establish the initial database.}  
Several seed graphs are randomly generated, with their normalized graph densities spreading  over $[0,1]$.  
Considering the small size of these graphs, we can calculate the quasi-optimal  parameters $(\gamma,\beta)$~\cite{brandao,guerreschi2017practical} as shown in Methods Section C.
These parameters provide potentially quasi-optimal parameters for  new graphs. 
This feature is particularly appealing to relatively larger target graphs.
The database can then be established, based on the pairs  $(n,D,\gamma,\beta)$.

\textbf{Substep 2: Develop the mapping table.} 
The mapping table is designed for transferring quasi-optimal parameters from seed graphs to target graphs with the same circuit layer number, $p$.
For creating the mapping table, several target graphs are also randomly generated, with normalized graph densities spread over $[0,1]$. 
With the parameters obtained from the seed graphs, QAOA calculation is performed for each target graph to get the cost function value $C(\ket{Z})$. 
Assume there are $\mathcal{N}$ seed graphs with  $n_{\rm{s}}$ vertices and $\mathcal{M}$  target graphs with  $n_{\rm{t}}$ vertices, the values of $C(\ket{Z})$  are then organized into a $\mathcal{N}\times \mathcal{M}$ matrix $M_{p}{(n_{\rm{s}} \mapsto n_{\rm{t}})}$, i.e., the mapping table. 
In the  table, each column is corresponding to one target graph and each row is corresponding to one seed graph. 
Note a mapping table only needs to be prepared once in advance for the same $p$, $n_{\rm{s}}$, $n_{\rm{t}}$.

\textbf{Substep 3: Transfer parameters to new graphs.} 
For a new graph with normalized graph density $D'$ and size $n'_{\rm{t}}$, several appropriate seed  graphs will be selected from the mapping table $M_{p}{(n_{\rm{s}} \mapsto n'_{\rm{t}})}$, whose size $n_{\rm{s}}$ needs to be equal or close to $n'_{\rm{t}}$.
Then, in the mapping table, one (or more) column, whose corresponding $D$  is equal or close to the new graph's $D'$, will be selected. 
Since each entry of this column is associated with a pair  $(n_{\rm{s}},\tilde{D},\gamma,\beta)$  of the seed graph, we can choose entries that are bigger than a threshold to get quasi-optimal parameters for the new graph. 
Specifically, based on the obtained entries, the  parameters in the pair corresponding to each entry will be identified and then transferred to the new graph. 
In this sense, we can use an interval $[\tilde{D}_{\rm{l}},\tilde{D}_{\rm{r}}]$ to summarize the identified pairs and denote the mapping as $D'\mapsto[\tilde{D}_{\rm{l}},\tilde{D}_{\rm{r}}]$, as shown in Fig.~\ref{fig_system}.
To improve the result's accuracy,  the layer number $p$ can be accordingly increased, with  parameters obtained by  above  transfer strategy.

The presented quasi-optimal parameter database is expendable, as mentioned in the Step $5$ of the data-driven QAOA. In the Substep $3$, we obtain the identified quasi-optimal parameters. 
For one thing, these parameters can be directly applied to the QAOA calculation of the new graph.
The new result can then be added to the current mapping table  $M_{p}{(n_{\rm{s}} \mapsto n'_{\rm{t}})}$ as a new entry, which is associated with the  pair 
$(n_{\rm{s}},\tilde{D},\gamma,\beta)$.
For another, these parameters can also be  decent initial guesses for further optimizing the parameters. Then based on the optimized result, a new pair $(n'_{\rm{t}},D',\gamma',\beta')$ can  be added to the database to provide potential parameters for new graphs, which is equivalent to the Substep 1.

\section{Numerical  Justification of the Parameter Transfer Strategy}
Since rigorously estimating the QAOA performace on generic weighted graphs is still an open question~\cite{barak2021classical}, 
we provide numerical examples to justify the effectiveness of the parameter transfer strategy.
We verify the efficacy of the transferred parameters from three aspects by comparing the approximation ratios with the ones obtained by QAOA using random parameters, QAOA using optimized parameters and GW algorithm.

For developing the mapping tables,
we randomly generate $9$ unweighted graphs with $n_{\rm{s1}}=10$ and $9$ weighted graphs with $n_{\rm{s2}}=24$ as seed graphs, as given in Extended  Table~\ref{tab_small_graphs_angles}. 
The $1710$ non-planar target graphs with $n_{\rm{t}}=24$ are also randomly generated, including $714$ unweighted graphs and $996$ weighted ones. 
The justifications involve $39,744$ QAOA expectation value calculations and at least
$16,146,548,640$ shots.
We apply two classical optimizers based on Newton and COBYLA methods~\cite{powell2007view} to get the mean approximation-ratio of the seed graphs and their corresponding quasi-optimal parameters. The values for  $p=1, 2, 3$ are summarized in Extended  Table~\ref{tab_small_graphs_angles}. 
These  parameters $(\gamma,\beta)$ provide the  initial data for the expandable  database as introduced in  Fig.~\ref{fig_system}.  

According to the Substep 2, we develop $12$ mapping tables as examples, among which the $6$ mapping tables in Extended  Fig.~\ref{fig_angles_index} are for the unweighted seed graphs under both weighted and unweighted target graphs with $p=1,2,3$, respectively; the other $6$ mapping tables in Extended  Fig.~\ref{fig_angles_index_24nodes} are for the weighted seed graphs. 

\subsection{Comparison with Using Random Parameters}
We compare the QAOA results from the transferred parameters and from random parameters to verify the parameter transfer strategy.
We apply  each seed graph's quasi-optimal parameters to the QAOA calculation for the $1,710$ target graphs, respectively.
Fig.~\ref{fig_meanValue_before_cancel_random} summarizes the mean approximation-ratios of QAOA in the unweighted and weighted graphs for $p=1,2,3$. In Fig.~\ref{fig_meanValue_before_cancel_random}, each black circle represents the mean approximation-ratio for a target graph with the identified parameters obtained from the $9$ groups parameters in  the mapping tables developed from unweighted $n_{\rm{s1}}=10$ seed graphs. These black circles show the high approximation-ratios, which are $0.8501,0.8911,0.9125$ in average for $p=1,2,3$, respectively.
Comparing these black circles with the pink circles showing the approximation ratios with random parameters, we can see that the parameter transfer strategy significantly improves the approximation ratio, especially for low density graphs, which verifies the effectiveness of the  transfer strategy. 
In addition, in Fig.~\ref{fig_meanValue_24nodes_before_cancel_random}, we also observe high approximation-ratio with parameters transferred from weighted $n_{\rm{s2}}=24$ seed graphs.

Meanwhile, we have two insights in both Fig.~\ref{fig_meanValue_before_cancel_random} and ~\ref{fig_meanValue_24nodes_before_cancel_random}. First, the significant improvement in low density graph is particularly appealing to cyber-physical systems, which usually have low densities. Second, it is challenging for the random parameters method to efficiently handle barren plateaus, while our method can address this issue by providing quasi-optimal initial guesses~\cite{wang2021noise}.
Further explanations for the effectiveness of the transfer strategy are shown in Methods Section E.

\subsection{Comparison with Using Optimized Parameters}

We verify that the transferred parameter can provide warm starting  for QAOA.

Fig.~\ref{fig_meanValue_iteration} compares the mean approximation-ratio  of QAOA using the transferred and unfavorable  parameters in the mapping table with and without further optimization.

On the one hand, the transferred parameters can be decent initial guesses.  Fig.~\ref{fig_meanValue_iteration} shows that after further optimizing the transferred parameters (blue scatters), the result  (orange scatters) has no significant  improvement. 
On the other hand, those transferred parameters also can be quasi-optimal  parameters.
Most of blue scatters without optimization are better than the purple scatters, which are optimized from yellow scatters with lots of computational effort.
There is a significant increment when comparing the transferred parameters with the worst parameters in the mapping table, as shown by the blue  and yellow scatters in Fig.~\ref{fig_meanValue_iteration}.
These comparisons validate the effectiveness of the parameter transfer strategy. 

In addition, the optimized parameters associated to the orange scatters in Fig.~\ref{fig_meanValue_iteration} can be adopted to expand the database.

\subsection{Comparison with the GW Algorithm}

\begin{figure}[ht]
	\centering{\includegraphics[width=0.5\textwidth]{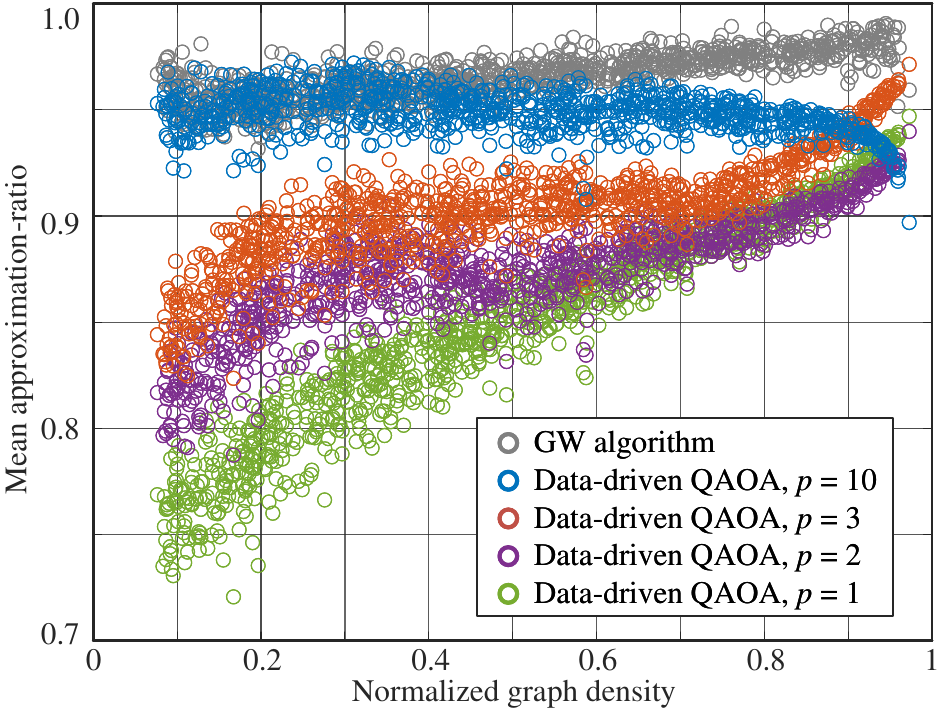}}
	\caption{The comparison of the GW algorithm and our data-driven QAOA with different layer numbers without parameter optimization.}
	\label{fig_QAOA_GW_comparison_envelope}
\end{figure}

For further verifying our strategy can provide the promising results, we compare the results of using the GW algorithm with the ones via the data-driven QAOA. $p=1,2,3,10$ are adopted as examples. 
In Fig.~\ref{fig_QAOA_GW_comparison_envelope}, when $p$ increases, the overall performance of transferred parameters increases. when $p=10$, the transferred parameters without any optimization have better mean approximation-ratio  than GW algorithm in the $113$ graphs out of total $996$ graphs. 
It is encouraging that, without any parameter optimization, the data-driven QAOA is competitive with GW algorithm.
We expect that proper optimization and larger $p$ could improve the approximation ratio further.

The drop trend of approximation ratio with $p=10$ in the large graph density regime, as shown in Fig.~\ref{fig_QAOA_GW_comparison_envelope}, is due to the overfitting on the seed graphs with $D=0.9111$ and $D=1$.
The $20$ parameters in the $10$-layer QAOA circuits could be excessive to be justified for some $10$-vertex seed graphs.
Notably, $p = \mathcal{O}(\log(n))$ could be sufficient to obtain high approximation-ratio solutions~\cite{barak2021classical}. 
The overfitting issue could be resolved in large seed graphs with $n$ vertexes and $p=\mathcal{O}(\log(n))$.

\begin{table*}
\centering
\caption{ The QAOA Results in Four Test Graphs with Different Normalized Graph Densities}
\label{tab_experiments}
\begin{tabular}{@{}cccccccccccccc@{}}
\toprule
\multirow{2}{*}{No.} &
\multirow{2}{*}{\begin{tabular}[c]{@{}c@{}}seed graph \\$\Tilde{D}$ \end{tabular}} &
  \multicolumn{3}{c}{$C(\ket{Z})(D=0.0525)$} &
  \multicolumn{3}{c}{ $C(\ket{Z})(D=0.1053)$} &
  \multicolumn{3}{c}{ $C(\ket{Z})(D=0.1143)$} &
  \multicolumn{3}{c}{ $C(\ket{Z})(D=0.3280)$} \\ \cmidrule(l){3-5}\cmidrule(l){6-8}\cmidrule(l){9-11}\cmidrule(l){12-14}
  &  & $p=1$   & $p=2$   & $p=3$   & $p=1$   &$p=2$   & $p=3$  & $p=1$   &$p=2$   & $p=3$   & $p=1$   &$p=2$   & $p=3$ \\ \midrule
1 & 0.2667 & \textbf{0.6865} & \textbf{0.7393} & \textbf{0.7724} & \textbf{0.7840} & \textbf{0.8180} & \textbf{0.8315} & \textbf{0.7828} & \textbf{0.8260} & \textbf{0.8495} & 0.8159 & 0.7581 & 0.7475 \\ \midrule
2 & 0.5333 & 0.6316 & 0.7059 & 0.7445 & 0.7577 & 0.7985 & 0.8216 & 0.7590 & 0.8150 & 0.8448 & 0.8448 & 0.8878 & 0.9107 \\ \midrule
3 & 0.6444 & 0.6135 & 0.6987 & 0.7368 & 0.7459 & 0.7941 & 0.8163 & 0.7431 & 0.8098 & 0.8384 & \textbf{0.8466} & 0.8940 & 0.9177 \\ \midrule
4 & 0.7333 & 0.6010 & 0.6991 & 0.7396 & 0.7373 & 0.7958 & 0.8176 & 0.7308 & 0.8110 & 0.8378 & 0.8450 & 0.8958 & \textbf{0.9189} \\ \midrule
5 & 0.8000 & 0.5920 & 0.7028 & 0.7460 & 0.7309 & 0.7965 & 0.8206 & 0.7216 & 0.8115 & 0.8363 & 0.8424 & 0.8963 & 0.9138 \\ \midrule
6 & 0.8667 & 0.5848 & 0.6892 & 0.7427 & 0.7258 & 0.7882 & 0.8209 & 0.7138 & 0.8027 & 0.8376 & 0.8393 & \textbf{0.8965} & 0.9146 \\ \midrule
7 & 0.9111 & 0.5793 & 0.6596 & 0.7259 & 0.7218 & 0.7769 & 0.8133 & 0.7077 & 0.7735 & 0.8306 & 0.8365 & 0.8669 & 0.9115 \\ \midrule
8 & 0.9556 & 0.5749 & 0.6448 & 0.6708 & 0.7183 & 0.7676 & 0.7876 & 0.7024 & 0.7537 & 0.7720 & 0.8337 & 0.8516 & 0.8607 \\ \midrule
9 & 1.0000 & 0.5710 & 0.4736 & 0.6386 & 0.7155 & 0.6593 & 0.7648 & 0.6979 & 0.6244 & 0.7454 & 0.8310 & 0.7617 & 0.8482 \\ \bottomrule
\end{tabular}
 \begin{tablenotes}
\footnotesize
\item[*] The bold expectation is the best value in the column.
\end{tablenotes}
\end{table*}
\section{Numerical Examples of Data-Driven QAOA on Cyber-Physical  Power Systems}

\begin{figure*}[!t]
	\centering{\includegraphics[width=\textwidth]{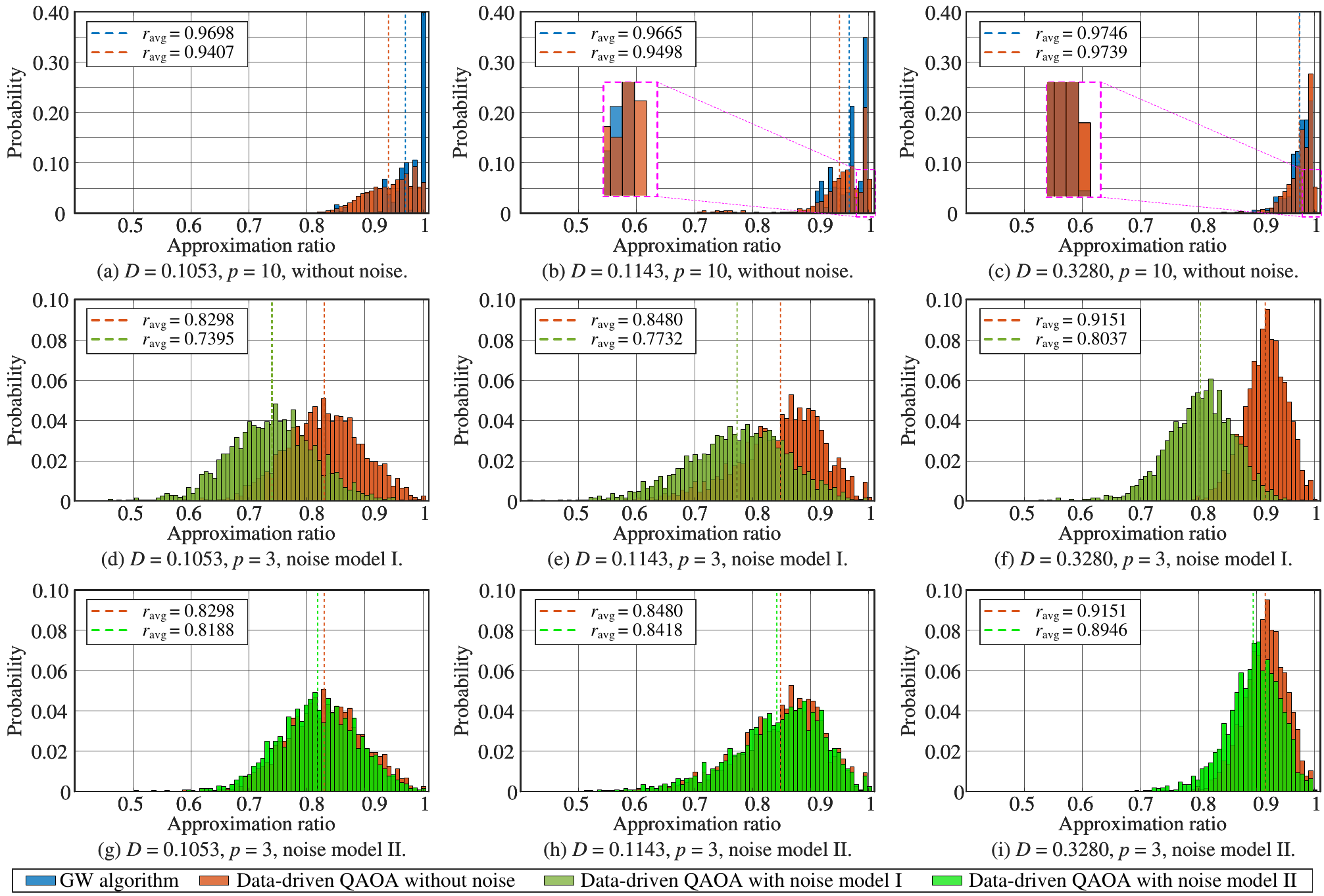}}
	\caption{The comparison of the approximation-ratio distributions between the GW algorithm and the date-driven QAOA with different layer numbers. Noise model I is $0.1\%$ depolarizing error on  single-qubit gates and $1\%$ depolarizing error on two-qubits gates; Noise model II is $0.01\%$ depolarizing error on single-qubit gates and $0.1\%$ depolarizing error on two-qubits gates.
	}
	\label{fig_QAOA_GW_comparison}
\end{figure*}

We test and verify the data-driven QAOA on a typical cyber-physical power system. The physical system is a modified  IEEE 24-bus system~\cite{ordoudis2016updated}, as given in Extended  Fig.~\ref{fig_IEEE24}. Eleven DERs are integrated into the system. Considering the output fluctuations of DERs, the normalized graph density will correspondingly change over time. So, two operational scenarios with  normalized graphs densities $D=0.0525$ and $D=0.1053$ are given as examples for the test. 
The communication network also has  $24$ vertices. Considering the dynamic data traffic in the network, two scenarios are  considered as examples with $D=0.1143$ and $D=0.3280$, respectively.
We provide the results from the following two aspects.

\subsection{Test without the Depolarizing Noise}
We carry out the test according to the steps  given in  Fig.~\ref{fig_system}.
Based on the power flow calculation of the physical system or the data traffic measurement of the cyber layer, four normalized weighted graphs can be obtained. 
With the normalized graph densities, quasi-optimal  parameters can be identified through the mapping table in Extended  Fig.~\ref{fig_angles_index} for the QAOA calculation. 
To provide a comparison, Table~\ref{tab_experiments} summarizes  the mean approximation-ratios  with all the  parameters in  the mapping table when $p=1,2,3$ for the four graphs.
The highlighted results emphasize that the best results  based on  the mapping table  can be obtained with 
the transferred parameters.
Thus, it justifies the effectiveness of the parameter transfer strategy. 

We also compare the data-driven QAOA's results with the GW algorithm. 
Fig.~\ref{fig_QAOA_GW_comparison} (a), (b), and (c) show the approximation-ratio distributions of different normalized  graph densities when $p=10$, with the following  findings.
First, the results show that the approximation means  
are very close to those of the GW algorithm. 
More importantly, Fig.~\ref{fig_QAOA_GW_comparison} (b) and (c) show that the data-driven QAOA's results are better than the GW algorithm, as there is at least ten times higher probability for the data-driven QAOA method than the GW algorithm to get the highest approximation ratio, as shown in the zoom-in details. In  practice, we usually use the highest cut value as an approximate solution instead of the mean approximation-ratio. The data-driven QAOA can be better than the GW algorithm.
Note these parameters are transferred from the mapping tables without any further optimization. Hence, according to Section VI B, when these parameters are used as initial guesses for further optimizing them,  the better mean approximation-ratio are $0.9569,0.9499,0.9751$ for the cases in Fig.~\ref{fig_QAOA_GW_comparison} (a), (b), and (c), respectively.
Second, 
Table~\ref{tab_experiments} shows that the mean approximation-ratio will increase as $p$ increases; and thus, a relatively larger $p$ is recommended for practical applications.

\subsection{Test with the Depolarizing Noise}
To verify the practicability of data-driven QAOA, we introduce the depolarizing noise on quantum gates to  simulate the realistic noise on quantum simulators~\cite{nielsen2011quantum}. 
Two noise models are considered. The noise model I is with  $0.1\%$ depolarizing error on single-qubit gates and $1\%$ depolarizing error on two-qubits gates, which is presently achievable. 
The noise model II is with $0.01\%$ depolarizing error on  single-qubit gates and $0.1\%$ depolarizing error on two-qubits gates, which is achievable in the near term.  

We carry out the numerical noise experiments on the test graphs. Fig.~\ref{fig_QAOA_GW_comparison} (d)-(i) show the examples of three graphs  with $D=0.1053$, $D=0.1143$, and $D=0.3280$, under the transferred parameters when  $p=3$.
By comparing the approximation-ratio distributions and means  between the results with and without noise,  we can see that the mean  approximation-ratios drop, with noise model I, as given in Fig.~\ref{fig_QAOA_GW_comparison} (d), (e), and (f). 
While with the smaller noise, the reduction of mean approximate-ratio is negligible, as shown in Fig.~\ref{fig_QAOA_GW_comparison} (g), (h), and (i).
Therefore, it is feasible to run the data-driven QAOA on a NISQ quantum processor and address the Max-Cut problem in the practical power system in the near term.

\section{Conclusions}
We present a data-driven QAOA to efficiently search for the maximum power or data sections in DER dominant cyber-physical power systems by leveraging quantum advantage. 
The parameter transfer strategy is designed to provide quasi-optimal parameters from seed graphs to  target graphs. It addresses the challenge of obtaining the critical parameters in QAOA; and thus, highly improving the efficacy and efficiency of QAOA.
In the transfer strategy, normalized graph density is utilized to bridge the seed and target graphs for  developing an extendable mapping table. 
We have verified the transfer strategy by comparing our approximation ratios with those obtained by QAOA using random parameters, QAOA using optimized parameters and GW algorithm.
The parameter transferability has also been verified from two perspectives, namely between unweighted and weighted graphs and  between small scale  and large scale  graphs as well as  graphs with the same size.
We simulate the presented method in a modified IEEE 24-bus  system and  demonstrated its  effectiveness in finding the maximum sections with and without depolarizing noise.
The presented method showcases the new computation of power systems  when meeting quantum technology. 
As a promising early candidates for achieving quantum advantage on NISQ systems, it can also be extended to address challenging issues in other complex engineered systems and eventually evolve into a formal quantum methodology.

\section{Methods}
\subsection{Adiabatic Approximation with QAOA}
According to the adiabatic evolution theorem~\cite{farhi2000quantum}, during the time interval $[0,T]$, we can slowly change the system's Hamiltonian from $H_{\rm{B}}$ to $H_{\rm{C}}$ and obtain the maximum energy state of $H_{\rm{C}}$ with high probability~\cite{farhi2014quantum}.
The changing process is exampled in (\ref{eq_Hamiltonian_evolution}).
\begin{eqnarray} 
\label{eq_Hamiltonian_evolution}
{H}\left(t\right) &=&\left[1-s\left(t\right)\right]{H}_{\rm{B}}+s\left(t\right){H}_{\rm{C}},
\end{eqnarray}
where $s\left(t\right)$ is a smooth function, $s\left(0\right)=0$ and $s\left(T\right)=1$. 
We then use Trotterization technique to emulate the evolution process~\cite{wu2002polynomial}.

We discretize the total time interval $[0,T]$ into intervals $[j\Delta t, (j+1)\Delta  t]$  with small enough $\Delta t$. Over the $j^{\text{th}}$ interval, the Hamiltonian is approximately constant, i.e., $H(t)=H((j+1)\Delta  t)$.  Therefore,  the total time evolution operator $U(T,0)$ can be approximately discretized into $2p$ implementable operators with constant Hamiltonian~\cite{wu2002polynomial}, as written in (\ref{eq_time_evolution_operator}).
The approximation will improve as $\Delta t$ gets smaller or, equivalently, as $p$ gets bigger.

\begin{equation} \label{eq_time_evolution_operator} 
%U_j
\begin{aligned}
U\left(T,0\right)&=U(T,T-\Delta t)U(T-\Delta t,T-2\Delta t) \cdots U(\Delta t,0)\\
&=\prod_{j=0}^{p-1}U((j+1)\Delta t, j\Delta t)%U((j+1)\Delta t,j \Delta t)
\approx\prod_{j=1}^{p}e^{-i{H}\left({j}\Delta t\right)\Delta t},
\end{aligned}
\end{equation}
\noindent
where $U((j+1)\Delta t, j\Delta t)$ represents the time evolution from  $ j\Delta t$  to  $(j+1)\Delta t$. 
Inserting (\ref{eq_Hamiltonian_evolution}) to (\ref{eq_time_evolution_operator}) and using  $e^{i(A_1+A_2)x} = e^{iA_1x}e^{iA_2x}+\mathcal{O}(x^2)$, the time evolution operator can be expressed as,
\begin{equation} \label{eq_time_evolution_operator_1} 
\begin{aligned}
U\left(T,0\right)&\approx\prod_{j=1}^{p}e^{-i\left[\left(1-s\left(j\Delta t\right)\right){H}_{\rm{B}}+s\left(j\Delta t\right){H}_{\rm{C}}\right]\Delta t}\\
&\approx\prod_{j=1}^{p}{e^{-i\left(1-s\left(j\Delta t\right)\right){H}_{\rm{B}}\Delta t}e^{-is\left(j\Delta t\right){H}_{\rm{C}}\Delta t}}+O(\Delta t^{2})\\
&\approx\prod_{j=1}^{p}{U_{\rm{B}}^{\left(j\right)}U_{\rm{C}}^{\left(j\right)}},
\end{aligned}
\end{equation}

\noindent
where $U_{\rm{B}}^{\left(j\right)}$ and $U_{\rm{C}}^{\left(j\right)}$ are the time evolution operators, evolving the system under the Hamiltonian $H_{\rm{B}}$ for the time period of $\beta_j=(1-s(j\Delta t))\Delta t$ and the Hamiltonian $H_{\rm{C}}$ for the time period of $\gamma _j=s(j\Delta t)\Delta t$, respectively, as defined in (\ref{eq_unitary_operator}).

\begin{equation} \label{eq_unitary_operator} 
\left \{
\begin{array}{lr}
U_{\rm{C}}^{\left(j\right)}=e^{-is\left(j\Delta t\right){H}_{\rm{C}}\Delta t}=e^{-i\gamma_j{H}_{\rm{C}}}&\\
U_{\rm{B}}^{\left(j\right)}=e^{-i\left[1-s\left(j\Delta t\right)\right]{H}_{\rm{B}}\Delta t}=e^{-i{\beta_j}{H}_{\rm{B}}}&
\end{array}
\right.
\end{equation}

In the evolution, $\ket{\varphi}$ represents the quantum state $\ket{Z}$ in section III.
Through applying $U_{\rm{B}}^{\left(j\right)}$ and $U_{\rm{C}}^{\left(j\right)}$ to the initial state $\ket{\varphi(0)}={\ket{+}}^{\otimes n}$ alternately, we can compute the final state $\ket{\varphi(T)}$ in (\ref{eq_final_state_UBUC}), which is expected to lead a high  $C(\ket{\varphi(T)})$ and collapse to maximum energy state after measurement.

 \begin{equation} \label{eq_final_state_UBUC}
 \begin{aligned}
{\ket{\varphi\left(T,{{\gamma}},{{\beta}}\right)}= \prod_{k=1}^{p}U_{\rm{B}}^{\left(j\right)}U_{\rm{C}}^{\left(j\right)}\ket{\varphi\left(0\right)}},
\end{aligned}
\end{equation}
where  ${{\gamma}}=\left(\gamma_1,\gamma_2,\ldots,\gamma_{p}\right)$ and ${{\beta}}=\left(\beta_1,\beta_2,\ldots,\beta_{p}\right)$  need to be optimized, which requires expensive  computational  effort.

\subsection{Measurement Outcomes for the Cost Function Value}

Quantum computers perform calculations based on the probability distribution of quantum states.
In QAOA, we obtain the cost function value  in  (\ref{eq_quantum_cost_function})  by sampling the quantum states, where  $|\alpha_k|^2$ is the probability that the final state $\ket{\varphi}$ collapses on the computational basis $\ket{Z_k}$, as explained below.

First, we construct the quantum circuit of QAOA for the target graphs. In our study, the circuit is built in the Qiskit simulator~\cite{Qiskit}. The quantum circuit  prepares the initial maximum energy state and computes the final state in (\ref{eq_final_state_UBUC}) by using the quantum operators $U_{\rm{B}}^{\left(j\right)}$ and $U_{\rm{C}}^{\left(j\right)}$, whose implementations are illustrated in Fig.~\ref{fig_system} and also shown in (\ref{eqn_implement_HB}) and (\ref{eqn_implement_HC}), respectively.

\begin{equation}
\begin{aligned}
\label{eqn_implement_HB}
%=e^{-i\beta_kH_{\rm{B}}}
e^{-i{\beta_k}{H}_{\rm{B}}}=e^{-i\beta_k\sum_{j=1}^n \sigma_j^x}=  \prod_{j=1}^{n}e^{-i\beta_k \sigma_j^x}=\prod_{j=1}^{n}R_{\rm{X}}^{(j)}(2\beta_k),\\
% \bigotimes_{j=1}^n R_X(2\beta_k)
\end{aligned}
\end{equation}
\begin{equation}
\begin{aligned}
\label{eqn_implement_HC}
e^{-i\gamma_k{H}_{\rm{C}}}=e^{-i\gamma_k\sum_{}^{} w_{ij}\frac{I-\sigma_i^z \sigma_j^z}{2}}= \prod_{ {\left \langle i,j\right \rangle} \in E}^{}R_{\rm{ZZ}}^{\left \langle i,j\right \rangle}(-\gamma_k w_{ij}),
\end{aligned}
\end{equation}
\noindent
where
$R_{\rm{X}}^{(j)}$ means only applying $R_{\rm{X}}$ gate to the $j^{\text{th}}$ qubit without changing other qubits; and $R_{\rm{ZZ}}^{\left \langle i,j\right \rangle}$ means only applying $R_{\rm{ZZ}}$ gate to the $i^{\text{th}}$ and  $j^{\text{th}}$ qubits.

Second, we run the quantum circuit $N_{\rm{shot}}$ times and measure the final state for the probability distribution. Suppose the final state collapses on the $\ket{Z_k}$ with $N_k$ times, the approximation of $|\alpha_k|^2$ is $|\tilde{\alpha}_k|^2=N_k/N_{\rm{shot}}$. The approximation will improve as $N_{\rm{shot}}$ gets bigger. 
In this study, to get an accurate probability distribution and mean approximation-ratio, we perform $N_{\rm{shot}}=2^{19}$ to approximate the distribution coefficients $|\alpha_k|^2$. In practice, $2,048$ shots works well and is recommended.

Third, we calculate  the cost function $C(\ket{\varphi})$ as shown in (\ref{eq_quantum_cost_function}), which is the weighted summation of $C(\ket{Z_k})$ with the non-zero coefficients $|\tilde{\alpha}_k|^2$.
Since the standard deviation of $C(\ket{\varphi})$ is in the order of $\sqrt{m}$~\cite{farhi2014quantum},  $N_{\rm{shot}}$ is in the polynomial order. So, it is efficient to compute  $C(\ket{Z_k})$ with non-zero coefficients $|\tilde{\alpha}_k|^2$.
Based on the calculation of $C(\ket{Z_k})$ with  non-zero coefficients $|\tilde{\alpha}_k|^2$, we select the $\ket{Z_{\rm{opt}}}$ with the maximal cut value as the final solution $Z_{\rm{opt}}$. 

\subsection{ Optimization of Parameters}

This subsection explains the parameter optimization involved in  three perspectives of Section VI and Section VII, where we need to optimize the parameters for high cost function values in seed graphs with $n_{\rm{s}}=10$ and in target graphs with $n_{\rm{t}}=24$, when $p=1,2,3,10$. The three perspectives are introduced below.

First, we get the optimal parameters for the seed graphs with $n_{\rm{s}}=10$ when $p=1,2,3$ by classical optimization method.
In our study, the Newton method is used to get the exact cost function values. 
Considering the non-convex landscapes of the cost function, we adopt multiple initial guesses for $(\gamma,\beta)$ within  ${[0,2\pi]}^p\times{[0,\pi]}^p$. The number of initial guesses is designed in the polynomial order of $n$ and $m$, which is proved to be sufficient for obtaining the optimal parameters~\cite{farhi2014quantum}. 

Second, we get the quasi-optimal parameters for the seed graphs with $n_{\rm{s}}=10$ when $p=10$ by  FOURIER heuristic strategy~\cite{zhou2020quantum}.
In the FOURIER strategy, the time complexity of obtaining quasi-parameters is reduced  into $\mathcal{O}(\text{poly}(p))$ to avoid computational burden~\cite{zhou2020quantum}, thus the parameters with high $p$ can be obtained efficiently.

Third,  after getting the transferred parameters, we further  optimize them for verifying the efficacy  of  the  transferred  parameters. 
Specifically, we use the COBYLA method~\cite{powell2007view}  to further optimize the transferred parameters due to the fluctuations of the cost function values. 
As mentioned in Methods Section B, the quantum computer estimates the cost function values by sampling  copies of output quantum state, which results in the fluctuations of the cost function values.
The COBYLA method is used to address the optimization with this issue, with the optimized results given in Fig.~\ref{fig_meanValue_iteration} and Extended  Table~\ref{tab_small_graphs_angles}.
The COBYLA method is also used to carry out the  optimization of transferred parameters for the test cases without the depolarizing noise.
Note that some certain gradient-based methods might also be able to find the quasi-optimal parameters with fluctuating cost function value, where the inaccurate gradient estimation may cause the escape from a local maximum and allow the converge towards a better one~\cite{guerreschi2017practical}.

\subsection{ Test Graphs Preparation}
Without losing generality, the test graphs are randomly generated, as introduced below.
First, we randomly generate adjacency  matrices. Second,  we set the entries to be zero with different probability to ensure the densities of the test graphs spread over $[0,1]$. Third, since there exist a polynomial algorithm for Max-Cut problem for the planar graph~\cite{shih1990unifying}, we check the generated graphs' planar property by  Kuratowski's Theorem, such that all of the test graphs are not planar. 
Finally, we generate $11,840$ graphs, sort them by normalized graph densities, and then uniformly  pick out $1710$ graphs as test graphs. 

\subsection{ Findings for Transfer Principles}

Here we show two important findings in Fig.~\ref{fig_meanValue_before_cancel_random} to further explain the transfer principles.

First, the mean approximation-ratio of QAOA for the target graphs are correlated to a Lipschitz continuous curve  with respect to their normalized graph densities, although these target graphs are randomly generated.
It is justified by a scan window with the size $0.1$, as given in Fig.~\ref{fig_meanValue_before_cancel_random}. The window shows that the upper limit of the standard deviations of the scatters with the same parameters is $0.057$, which further indicates the scatters approximately follow a curve. 
It also indicates the normalized graph density is a effective metric.
According to these curves, we can directly estimate the mean approximation-ratio of new graphs with  parameters in the database. 
Thereafter the parameters with outstanding performance can be quickly identified for the QAOA circuit, avoiding the high computing effort.

Second, the parameters of seed graphs with low density perform better in target graphs with  low density than in the ones with high  density, and vice versa.
This property is also uncovered in the mapping tables, where the yellow area denoting the high approximation-ratio will increase as $D$ increases. Specifically, when the sizes of the seed and target graphs are very close, the yellow area will be around the diagonal line as shown in Extended  Fig.~\ref{fig_angles_index_24nodes}; while, when the size of the target graph is much bigger than that of the seed graphs, the area will be above the diagonal line as shown in Extended Fig. 	\ref{fig_angles_index}. With this property, the quasi-optimal parameters can be effectively identified. 

\subsection{Modeling the  Cyber-Physical Power System}
We model the physical layer and then get multiple normalized weighted graphs through the power flow calculation when disturbances from DERs are considered. 
Power flow calculates the bus voltages  for a given load, generation, and network condition, based on which the line powers (weights) can be obtained. 
The power flow equations are given in (\ref{eqn_power_flow}).
\begin{equation}
\label{eqn_power_flow}
\begin{aligned}
% \dot{S}_i^{*}=
P_i-jQ_i=\dot{V}_i^*\sum_{k=1}^n{\dot{y}_{ik}\dot{V}_k},
\end{aligned}
\end{equation}
where $*$ denotes conjugate, $\dot{V}_i\in \mathbb{C}$ is the $i^{\text{th}}$ bus (vertex) voltage in the physical grid, $P_i,Q_i\in \mathbb{R}$ is the injection active  and reactive power of the $i^\text{th}$ bus, and $\dot{y}_{ik}\in \mathbb{C}$ is the admittance of the line between the  $i^\text{th}$ bus and $k^\text{th}$ bus.

After solving the power flow equations, we can obtain the  complex  power over each  line. In our study, the apparent power is used as the edge weight. 
Due to the  complex landscape of parameters~\cite{Jing2022interoperation}, the edge weight is then normalized. 
Thus, the modeling graph for the physical system is a normalized weighted  graph. 

The modeling graph of the cyber layer is based on the communication network data traffic that is flexible and random. 
In our study, we randomly generate $n_{\rm{t}}=24$ graphs to represent the communication network. In practical applications, we can monitor the data traffic to  set up the edge weights for the cyber graphs. 

\subsection{Depolarizing Noise Model}
For  demonstrating the potential of our method to be a promising candidate for achieving quantum advantage on NISQ systems, we introduce the depolarizing noise for the quantum gates in QAOA circuits. 
In our study, Qiskit~\cite{Qiskit} is used to simulate the depolarizing noise and investigate the influences. 
The simulator needs to calculate the density matrix after each quantum gate to include the noise model, which costs exponentially more calculation resources than the noiseless vector simulation.

\subsection{The Approximation-ratio  Distribution of GW Algorithm}
In  Fig.~\ref{fig_QAOA_GW_comparison} (a)-(c), we obtain the approximation-ratio distributions of the GW algorithm, as summarized below.

The GW algorithm relaxes the constraint of Max-Cut problem from discrete variables to the vectors on a unit sphere. The relaxed problem then becomes a semidefinite programming (SDP) problem. By solving the SDP problem, we obtain the optimal vector distribution. By randomly cutting the unit sphere into two parts, we correspondingly separate the vectors into two groups and obtain an approximate solution.
When we cut the sphere several times, there is a guarantee that we have at least $0.878$ expected approximation ratio.

Similarly to the shots in QAOA, with certain  cut times, the GW algorithm outputs a probability distribution with respect to the approximation ratio, e.g., the approximation-ratio distribution. 
In the theoretical research, we usually compare the expectation of the approximation-ratio distribution to evaluate the algorithm performance, while in the practical applications to cyber-physical systems, we can take  the maximum approximation-ratio as the final approximation solution.

\section{Acknowledgment}
Y.W. is primarily supported by the Office of the
Director of National Intelligence—Intelligence Advanced Research Projects Activity through ARO Contract No. W911NF-16- 1-0082 and DOE BES award de-sc0019449 (quantum algorithm analysis). The author would like to thank Dr. Rui Chao at Duke University for the detailed discussion on the QAOA algorithm. 

Recently, we became aware of a similar work by Shaydulin \emph{et al.}  about the transferability between weighted graphs~\cite{shaydulin2022parameter}, which was carried out independently.

\printbibliography

\appendices

\begin{tableExt*}[]
\centering
\caption{ The Parameters Performance of Seed Graphs with Different Layer Numbers $p$ and Vertex Size $n_{\rm{s}}$}
\begin{tabular}{@{}ccccccccc@{}}
\toprule
\multirow{2}{*}{\begin{tabular}[c]{@{}c@{}}seed \\ graph\end{tabular}} &
   \multicolumn{4}{c}{$C(\ket{Z})$ when $n_s=10$} &
  \multicolumn{4}{c}{$C(\ket{Z})$ when $n_s=24$} \\ \cmidrule(l){2-5} \cmidrule(l){6-9} 
  &$D$  & $p=1$     & $p=2$        &$p=3$  & $D$  & $p=1$     & $p=2$        &$p=3$    \\ \midrule
1 & 0.2667 & 0.7783  & 0.8727 & 0.9287 & 0.0831 & 0.7395 & 0.8117 & 0.8573 \\ \midrule
2 & 0.5333 & 0.8537  & 0.9193 & 0.9516 & 0.1901 & 0.7931 & 0.8526 & 0.8866 \\ \midrule
3 & 0.6444 & 0.8447  & 0.9009 & 0.9296 & 0.2869 & 0.8122 & 0.8681 & 0.8950 \\ \midrule
4 & 0.7333 & 0.8182  & 0.8601 & 0.8885 & 0.3527 & 0.8363 & 0.8872 & 0.9114 \\ \midrule
5 & 0.8000 & 0.8797  & 0.9206 & 0.9493 & 0.5163 & 0.8518 & 0.8974 & 0.9178 \\ \midrule
6 & 0.8667 & 0.8672  & 0.8943 & 0.9170 & 0.6256 & 0.8793 & 0.9168 & 0.9325 \\ \midrule
7 & 0.9111 & 0.9044  & 0.9223 & 0.9396 & 0.6756 & 0.8743 & 0.9105 & 0.9259 \\ \midrule
8 & 0.9556 & 0.9420  & 0.9553 & 0.9639 & 0.8316 & 0.9152 & 0.9261 & 0.9353 \\ \midrule
9 & 1.0000 & 0.9804  & 0.9977 & 0.9999 & 0.9608 & 0.9588 & 0.9588 & 0.9661 \\ \bottomrule
\end{tabular}

\label{tab_small_graphs_angles}
    \begin{tablenotes}
      \small
      \item We randomly generate several seed graphs, including $9$ unweighted $n_{\rm{s1}}=10$ graphs with ${D}$ spreading over $[0.2667,1]$, and $9$ weighted $n_{\rm{s2}}=24$ graphs with ${D}$ spreading over $[0.0831,0.9608]$. In each seed graph, we use the classical optimizers mentioned in Methods Section C to obtain the quasi-optimal parameters, which have better approximation ratio as the layer number $p$ increases.
    \end{tablenotes}
    
\end{tableExt*}

\begin{figureExt*}[!t]
	\centering{\includegraphics[width=\textwidth]{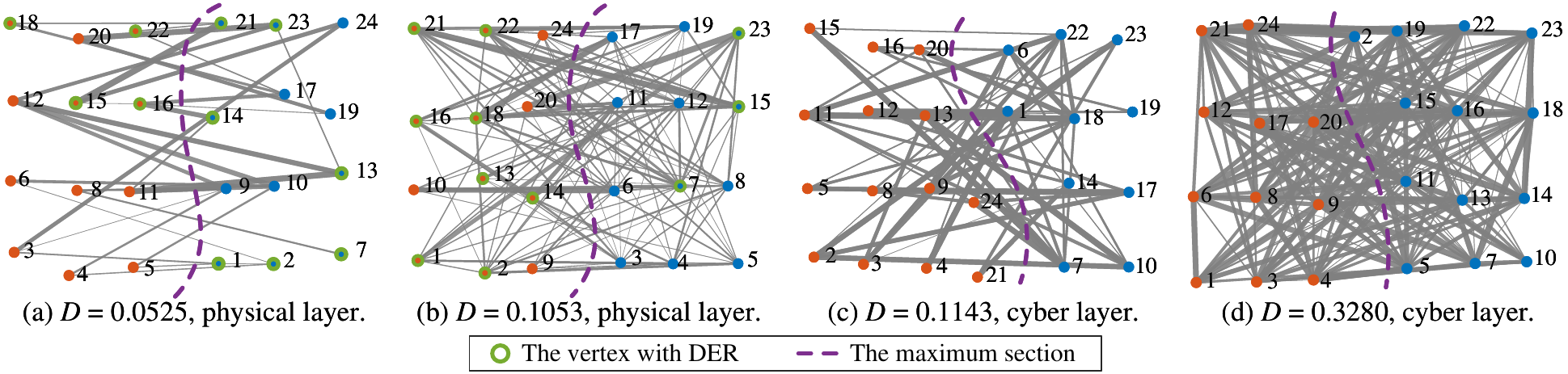}}
	\caption{The four scenarios of the test cyber-physical power system.
	}
	\label{fig_IEEE24}
\end{figureExt*}

\begin{figureExt*}[!t]
	\centering{\includegraphics[width=\textwidth]{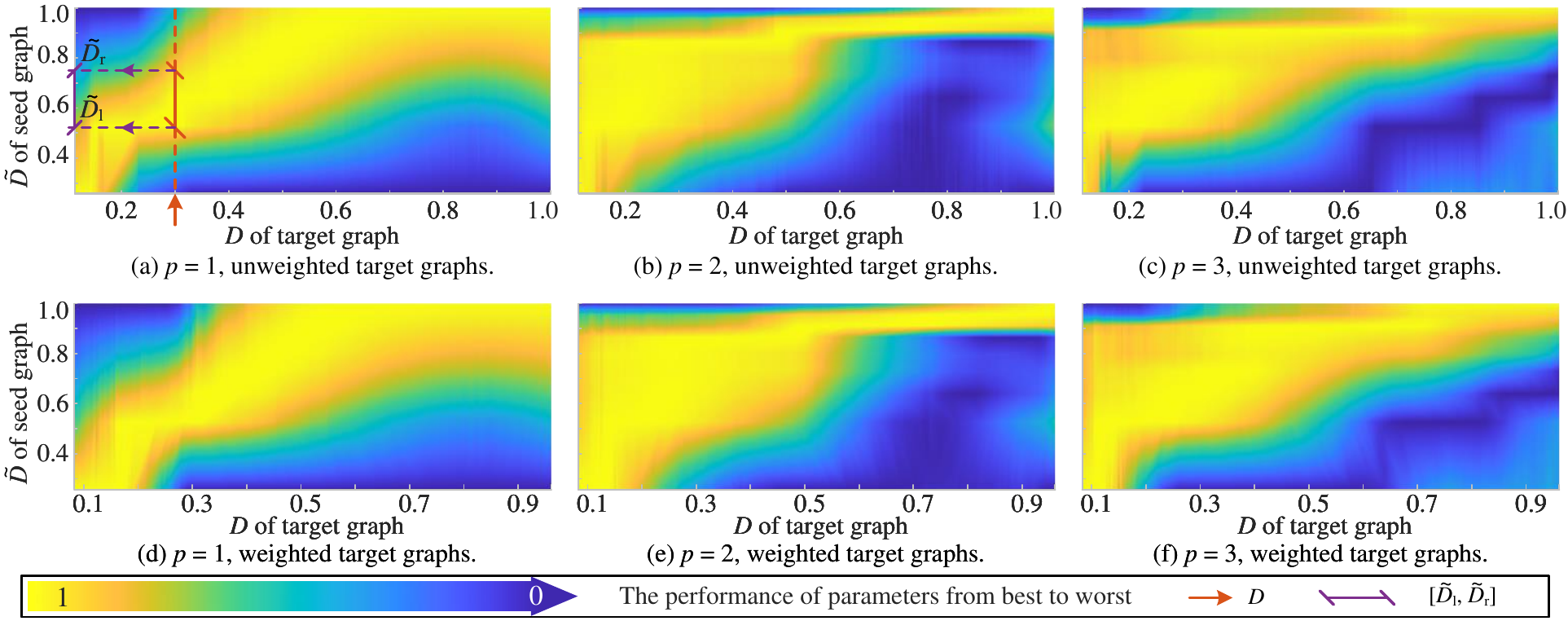}}
	\caption{The mapping table developed by the parameters from unweighted seed graphs with $n_{\rm{s1}}=10$. Each sub-figure is a mapping table for $D\mapsto[\tilde{D}_{\rm{l}},\tilde{D}_{\rm{r}}]$ in the parameter transfer module of the proposed data-driven QAOA in Fig. \ref{fig_system}. 
	This mapping $D\mapsto[\tilde{D}_{\rm{l}},\tilde{D}_{\rm{r}}]$ is performed  based on the entries in the mapping table. The entries are the scaling approximation ratios, which can be obtained  by applying the parameters from seed graphs to the QAOA for the target graphs, as shown in Fig.~\ref{fig_meanValue_before_cancel_random}.
	The procedure of getting $D\mapsto[\tilde{D}_{\rm{l}},\tilde{D}_{\rm{r}}]$ is that we first fix the horizontal axis as shown in the orange arrow according to $D$  and then find the interval $[\tilde{D}_{\rm{l}},\tilde{D}_{\rm{r}}]$ as shown in purple line according to the contour. After mapping, we identify the quasi-optimal parameters  from the seed graphs whose $\tilde{D}$ are within $[\tilde{D}_{\rm{l}},\tilde{D}_{\rm{r}}]$.}
	\label{fig_angles_index}
\end{figureExt*}

\begin{figureExt*}[!t]
	\centering{\includegraphics[width=\textwidth]{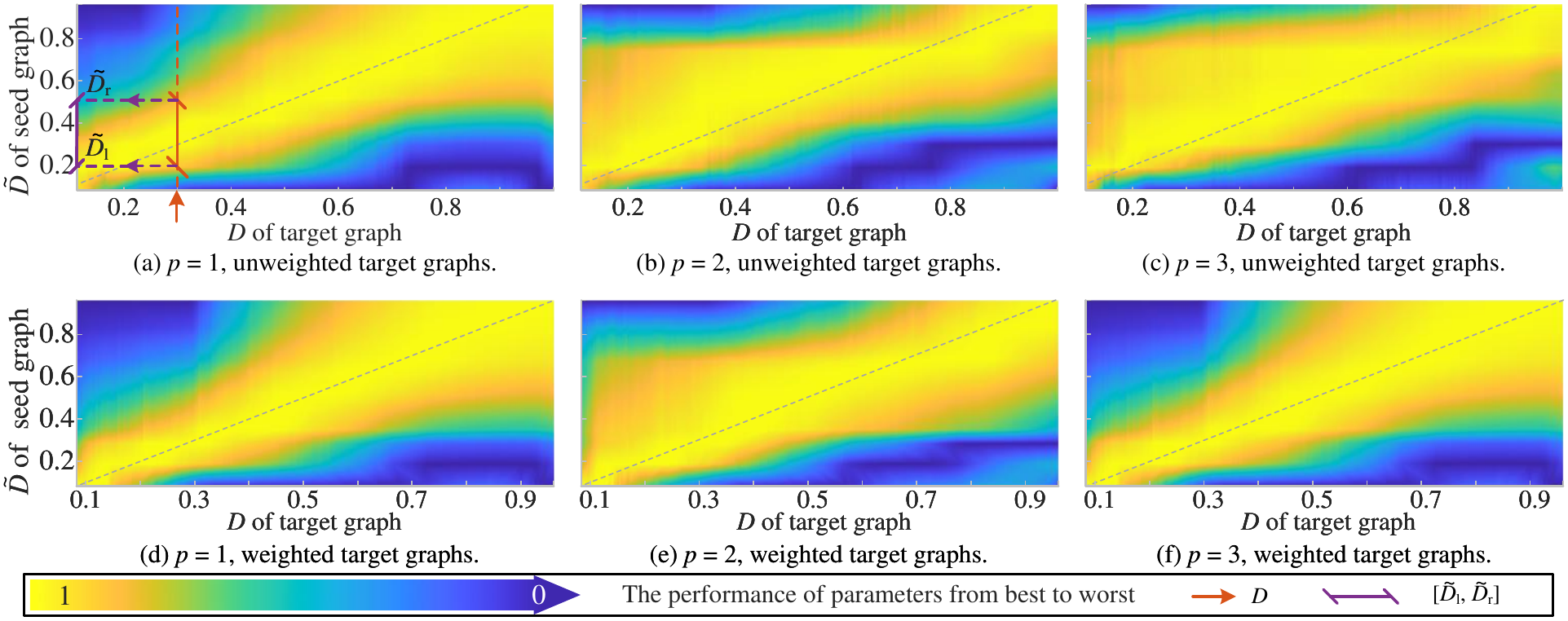}}
	\caption{The mapping table developed by the parameters from weighted seed graphs with $n_{\rm{s2}}=24$. Each sub-figure is a mapping table for $D\mapsto[\tilde{D}_{\rm{l}},\tilde{D}_{\rm{r}}]$ in the parameter transfer module of the proposed data-driven QAOA in Fig. \ref{fig_system}. The developing procedure of these mapping tables is the same with that of the mapping tables in Extended Fig. \ref{fig_angles_index}.  Note that since we identify the parameters  from $n_{\rm{s2}}=24$ seed graphs and apply them to the same size $n_{\rm{t}}=24$ target graphs, the maximal performances are along the diagonal dash line.}
	\label{fig_angles_index_24nodes}
\end{figureExt*}

\end{document}